# Adaptive Active-Passive Networked Multiagent Systems

Ehsan Arabi        Dimitra Panagou        Tansel Yucelen

*Abstract*— Active-passive multiagent systems consist of agents subject to inputs (active agents) and agents with no inputs (passive agents), where active and passive agent roles are considered to be interchangeable in order to capture a wide array of applications. A challenge in the control of active-passive multiagent systems is the presence of information exchange uncertainties that can yield to undesirable closed-loop system performance. Motivated by this standpoint, this paper proposes an adaptive control algorithm for this class of multiagent systems to suppress the negative effects of information exchange uncertainties. Specifically, by estimating these uncertainties, the proposed adaptive control architecture has the ability to recover the active-passive multiagent system performance in a distributed manner. As a result, the agents converge to a user-adjustable neighborhood of the average of the applied inputs to the active agents. The efficacy of the proposed control architecture is also validated from a human-robot collaboration perspective, where a human is visiting several task locations, and the multiagent system identifies these locations and move toward them as a coverage control problem.

## I. Introduction

There are considerable amount of studies in the literature focusing on dynamic consensus filters where a multiagent system is subject to observations of a process of interest (i.e., the agents can sense some exogenous inputs) [1–3]. Specifically, these studies consider that each agent/sensor is subject to an input (i.e., the agent can sense a process), where the goal of all agents is to converge to the average of the applied inputs. From a practical standpoint, however, a subset of the agents may not be subject to any inputs. To elucidate this point, consider a camera network as an example, where the goal is to locate the position of an intruder. In this example, only a few cameras in the network can see the intruder (i.e., they are subject to an input) at a given time, where it is required for all agents to converge to the average of the inputs of those few cameras for localization purposes. Hence, the results given in the aforementioned references cannot be utilized. A recently proposed approach called active-passive multiagent systems provides a remedy to this problem [4, 5]. This class of multiagent systems consist of agents subject to inputs (active agents) and agents with no inputs (passive agents). Specifically, in [4], an integral action-based distributed control algorithm is proposed, which requires agents to exchange their current state and integral state within each other, under the assumption that active and passive roles of agents are fixed in order to ensure all agents to converge to the average of the inputs of active agents. In [5], the requirements on integral state exchange and fixed active and passive roles are relaxed. Yet, how the presence of information exchange uncertainties can affect the convergence of all agents to the average of the inputs of active agents is largely unknown.

Furthermore, coverage is one of the fundamental objectives in multi-agent systems, either as static coverage [6–8] or dynamic coverage [9–12]. In particular, this problem addresses the optimal placement of sensors to cover a given region and has been extensively studied as the deployment of mobile sensing agents to the centroids of Voronoi cells in a Voronoi partition of a given domain based on a density map, that is usually assumed to be available through local measurements of the process of interest. However, obtaining a density map for converge control design for tracking a target such as in human-robot collaborative networks can be challenging, since the target information is only accessible for a subset of agents, resulting in an active-passive multiagent system as described above.

In this paper, we propose a distributed adaptive control architecture to address the aforementioned challenge in the active-passive multiagent systems in the presence of uncertainties in the communication signals. The proposed control architecture estimates the unknown information exchange uncertainties, and recovers the performance of active-passive multiagent system. Based on a Lyapunov stability analysis, we show that the proposed algorithm drives the agents to a user-adjustable neighborhood of the the average of the applied inputs in the face of these uncertainties. From a human-robot collaboration perspective, we then consider a converge control problem as a case study, where the agents are required to assist a human as it moves within a field. Specifically, considering the human as the only input to this active-passive multiagent system, the proposed adaptive control architecture estimates the position and velocity of the human. The agents then construct a density map and move toward the high density locations via coverage control.

## II. Mathematical Preliminaries

We start by introducing the notation used throughout this paper. $\mathbb{R}$, $\mathbb{R}^n$, and $\mathbb{R}^{n \times m}$ respectively denote the set of real numbers, the set of $n \times 1$ real column vectors, and the set of

Ehsan Arabi was a Postdoctoral Research Fellow at the Department of Aerospace Engineering, University of Michigan, when this work was conducted. He is currently a Research Engineer at Ford Research and Advanced Engineering, Ford Motor Company, Dearborn, MI 48121, USA (email: `earabi@ford.com`).

Dimitra Panagou is an Associate Professor with the Department of Aerospace Engineering, University of Michigan, Ann Arbor, MI 48109, USA (email: `dpanagou@umich.edu`).

Tansel Yucelen is an Associate Professor of the Department of Mechanical Engineering and the Director of the Laboratory for Autonomy, Control, Information, and Systems (LACIS) at University of South Florida, Tampa, FL 33620, USA (email: `yucelen@usf.edu`).

The first and second author would like to acknowledge the support of the Kahn Foundation under award number AWD011321.

$n \times m$ real matrices; $\mathbf{0}_n$ denotes the $n \times 1$ zero vector, $\mathbf{1}_n$ denotes the $n \times 1$ ones vector, $\mathrm{I}_n$ denotes the $n \times n$ identity matrix, and $\mathbb{R}_+$ (resp., $\overline{\mathbb{R}}_+$) and $\mathbb{R}_+^{n \times n}$ (resp., $\overline{\mathbb{R}}_+^{n \times n}$) denote the set of positive real numbers (resp., non-negative reals) and the set of $n \times n$ positive-definite (resp., positive semi-definite) real matrices. In addition, we use $(\cdot)^{\mathrm{T}}$ to denote the transpose, $(\cdot)^\dagger$ to denote generalized inverse, and $\lambda_{\min}(A)$ (resp., $\lambda_{\max}(A)$) to denote the minimum (resp., maximum) eigenvalue of the square matrix $A$.

Necessary notions from graph theory are next concisely overviewed (details in [13, 14]). Specifically, an undirected graph $\mathfrak{G}$ is defined by a set $\mathcal{V}_{\mathfrak{G}} = \{1, 2, \ldots, N\}$ of *nodes* and a set $\mathcal{E}_{\mathfrak{G}} \subset \mathcal{V}_{\mathfrak{G}} \times \mathcal{V}_{\mathfrak{G}}$ of *edges*. The *degree* of a node is given by the number of its neighbors. Letting $d_i$ denote the degree of node $i$, then the *degree matrix* of a graph $\mathfrak{G}$, denoted by $\mathcal{D}(\mathfrak{G}) \in \mathbb{R}^{N \times N}$, is given by $\mathcal{D}(\mathfrak{G}) \triangleq \mathrm{diag}\,[d]$, where $d = [d_1, \ldots, d_N]^{\mathrm{T}}$. We write $\mathcal{A}(\mathfrak{G}) \in \mathbb{R}^{N \times N}$ for the *adjacency matrix* and $\mathcal{B}(\mathfrak{G}) \in \mathbb{R}^{N \times M}$ for the *incidence matrix* of a graph $\mathfrak{G}$. The *graph Laplacian matrix*, $\mathcal{L}(\mathfrak{G}) \in \overline{\mathbb{R}}_+^{N \times N}$, is defined by $\mathcal{L}(\mathfrak{G}) \triangleq \mathcal{D}(\mathfrak{G}) - \mathcal{A}(\mathfrak{G})$ or equivalently $\mathcal{L}(\mathfrak{G}) = \mathcal{B}(\mathfrak{G})\mathcal{B}(\mathfrak{G})^{\mathrm{T}}$, where $\mathcal{L}(\mathfrak{G})\mathbf{1}_N = \mathbf{0}_N$ and $e^{\mathcal{L}(\mathfrak{G})}\mathbf{1}_N = \mathbf{1}_N$ hold. The following lemmas are needed for our main results, where we consider that a multiagent system is represented by a connected, undirected graph $\mathfrak{G}$.

**Lemma 1** ([15, Lemma 3.3]). *Let $K = diag(k)$, $k = [k_1, k_2, \ldots, k_N]^{\mathrm{T}}$, $k_i \in \overline{\mathbb{R}}_+$, $i = 1, 2, \ldots, N$, and assume that at least one element of $k$ is nonzero. Then, $\mathcal{F}(\mathfrak{G}) \triangleq \mathcal{L}(\mathfrak{G}) + K \in \mathbb{R}_+^{N \times N}$ and $\det(\mathcal{F}(\mathfrak{G})) \neq 0$ for the Laplacian of a connected, undirected graph $\mathfrak{G}$.*

**Lemma 2** ([16]). *For a connected, undirected graph, the Laplacian satisfies $\mathcal{L}(\mathfrak{G})\mathcal{L}^\dagger(\mathfrak{G}) = \mathrm{I}_N - \frac{1}{N}\mathbf{1}_N\mathbf{1}_N^{\mathrm{T}}$.*

Next, we state the definition of the projection operator. To this end, let $\Omega$ be a convex hypercube in $\mathbb{R}^n$ defined as $\Omega = \{\theta \in \mathbb{R}^n : (\theta_i^{\min} \leq \theta_i \leq \theta_i^{\max})_{i=1,2,\ldots,n}\}$, where $(\theta_i^{\min}, \theta_i^{\max})$ denote the minimum and maximum bounds for the $i$th component of the $n$-dimensional parameter vector $\theta$. Furthermore, let $\Omega_\nu$ be the second hypercube defined as $\Omega_\nu = \{\theta \in \mathbb{R}^n : (\theta_i^{\min} + \nu \leq \theta_i \leq \theta_i^{\max} - \nu)_{i=1,2,\ldots,n}\}$, where $\Omega_\nu \subset \Omega$ for a sufficiently small positive constant $\nu$.

**Definition 1** ([17, 18]). *For $y \in \mathbb{R}^n$, the projection operator* $\mathrm{Proj} : \mathbb{R}^n \times \mathbb{R}^n \to \mathbb{R}^n$ *is defined (componentwise) as* $\mathrm{Proj}(\theta, y) \triangleq ([\theta_i^{\max} - \theta_i]/\nu)y_i$ *when* $\theta_i > \theta_i^{\max} - \nu$ *and* $y_i > 0$, $\mathrm{Proj}(\theta, y) \triangleq ([\theta_i - \theta_i^{\min}]/\nu)y_i$ *when* $\theta_i < \theta_i^{\min} + \nu$ *and* $y_i < 0$, *and* $\mathrm{Proj}(\theta, y) \triangleq y_i$ *otherwise.*

Note that Definition 1 guarantees $(\theta - \theta^*)^{\mathrm{T}}(\mathrm{Proj}(\theta, y) - y) \leq 0$, $\theta^* \in \Omega_\nu$, where this inequality can be generalized to matrices using $\mathrm{Proj_m}(\Theta, Y) = (\mathrm{Proj}(\mathrm{col}_1(\Theta), \mathrm{col}_1(Y)), \ldots, \mathrm{Proj}(\mathrm{col}_m(\Theta), \mathrm{col}_m(Y)))$ with $\Theta \in \mathbb{R}^{n \times m}$, $Y \in \mathbb{R}^{n \times m}$, and $\mathrm{col}_i(\cdot)$ denoting $i$th column operator.

## III. OVERVIEW OF ACTIVE-PASSIVE MULTIAGENT SYSTEMS

For the main result of this paper presented in the following section, we here concisely overview the active-passive multiagent systems [5, 19, 20]. Consider $N$ agents exchanging information based on a connected, undirected graph $\mathfrak{G}$, where the dynamics of each agent satisfy

$$\dot{x}_i(t) = ax_i(t) + u_i(t), \quad x_i(0) = x_{i0}. \tag{1}$$

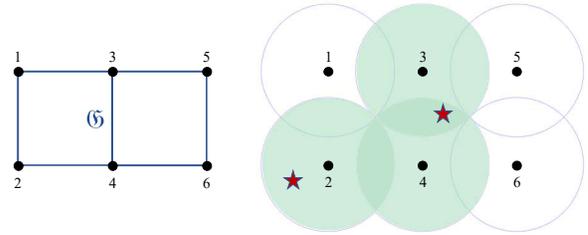

Fig. 1. Example of an active-passive multiagent system with six agents on a connected, undirected graph $\mathfrak{G}$. Agents 2, 3 and 4 are active at this time instance since they can sense the inputs (indicated by red stars).

Here, $x_i(t) \in \mathbb{R}$ is the state of agent $i$, $i = 1, \ldots, N$, $u_i(t) \in \mathbb{R}$ is the control input, and $a \in \mathbb{R}$, $a \neq 0$. The agents that can sense the exogenous inputs $c_h(t) \in \mathbb{R}$, $h = 1, \ldots, m$, $m \leq N$, are called *active* agents; otherwise, they are called *passive* agents. As the location of the exogenous input changes with respect to the agents, the active and passive roles of the agents changes (see Figure 1). The objective for the agents is to reach dynamic consensus to a user-adjustable neighborhood of the average of the applied inputs (i.e., the observable process by the active agents).

Consider the distributed control algorithm given by

$$\begin{aligned}
u_i(t) &= -k_0 x_i(t) - \alpha \Big[\sum_{i \sim j}\big(x_i(t) - x_j(t)\big) + \beta_i x_i(t)\Big] \\
&\quad + p_i(t) - \alpha \sum_{i \sim h} k_{ih}(t)\big(x_i(t) - c_h(t)\big), \tag{2}
\end{aligned}$$

$$\dot{p}_i(t) = -\gamma\Big[\sum_{i \sim j}\big(x_i(t) - x_j(t)\big) + \sigma p_i(t)\Big], \; p_i(0) = 0, \tag{3}$$

where $p_i(t) \in \mathbb{R}$ denotes the integral action of agent $i$. In (2) and (3), $\alpha, \gamma, \sigma \in \mathbb{R}_+$ are design parameters, $k_{ih}(t)$ is a smooth function that changes from 1 (at the location of the agent $i$) to 0 (at the sensing radius of the agent $i$), $k_0 \in \mathbb{R}$ is designed such that $a_0 \triangleq a - k_0 \leq 0$, and $\beta_i \in \overline{\mathbb{R}}_+$ is designed with at least one nonzero $\beta_i$, $i = 1, \ldots, N$. Define $\beta \triangleq \mathrm{diag}([\beta_1, \beta_2, \ldots, \beta_N]) \in \overline{\mathbb{R}}_+^{N \times N}$ and let $k_{1,i} \triangleq \sum_{i \sim h} k_{ih}(t) \in \overline{\mathbb{R}}_+$ denote the number of the inputs applied to active agent $i$. We now define $K_1(t) \triangleq \mathrm{diag}([k_{1,1}(t), k_{1,2}(t), \ldots, k_{1,N}(t)]) \in \overline{\mathbb{R}}_+^{N \times N}$ and

$$K_2(t) \triangleq \begin{bmatrix} k_{2,11}(t) & \cdots & k_{2,1N}(t) \\ \vdots & \ddots & \vdots \\ k_{2,N1}(t) & \cdots & k_{2,NN}(t) \end{bmatrix}, \tag{4}$$

such that $k_{1,i}(t) \triangleq \sum_{j=1}^{N} k_{2,ij}(t) \in \overline{\mathbb{R}}_+$ holds. In a compact form, one can now write (1), (2) and (3) as

$$\begin{aligned}
\dot{x}(t) &= a_0 x(t) - \alpha \mathcal{L}(\mathfrak{G})x(t) - \alpha\beta x(t) + p(t) - \alpha K_1(t)x(t) \\
&\quad + \alpha K_2(t)c(t), \tag{5}
\end{aligned}$$

$$\dot{p}(t) = -\gamma \mathcal{L}(\mathfrak{G})x(t) - \gamma\sigma p(t), \quad p(0) = \mathbf{0}_N, \tag{6}$$

where $x(t) \triangleq [x_1(t), \ldots, x_N(t)]^{\mathrm{T}} \in \mathbb{R}^N$, $p(t) \triangleq [p_1(t), \ldots,$

$p_N(t)]^\mathrm{T} \in \mathbb{R}^N$, and $c(t) \triangleq [c_1(t),\ldots,c_m(t),0,\ldots,0]^\mathrm{T} \in \mathbb{R}^N$. The average of the applied inputs to the active agents can be given by

$$\epsilon(t) \triangleq \frac{\mathbf{1}_N^\mathrm{T} K_2(t) c(t)}{\mathbf{1}_N^\mathrm{T} K_2(t) \mathbf{1}_N} \in \mathbb{R}. \tag{7}$$

The following assumption is utilized in studies concerning active-passive multiagent networks.

**Assumption 1.** *The exogenous inputs $c_h(t) \in \mathbb{R}$, $h = 1,\ldots,m$, $m \leq N$, and their derivatives are bounded.*

**Remark 1.** *It directly follows from Assumption 1 that: (i) there exist $\overline{c}, \overline{c}_d \in \mathbb{R}_+$ such that $\|c(t)\| \leq \overline{c}$ and $\|\dot{c}(t)\| \leq \overline{c}_d$, (ii) the average of the applied inputs and its derivative are bounded, that is, there exist $\overline{\epsilon}$ and $\overline{\epsilon}_d$ such that $\|\epsilon(t)\| \leq \overline{\epsilon}$ and $\|\dot{\epsilon}(t)\| \leq \overline{\epsilon}_d$.*

Define the error between the states of the agents and the average of the applied inputs as

$$\delta(t) \triangleq x(t) - \mathbf{1}_N \epsilon(t) \in \mathbb{R}^N. \tag{8}$$

Now let

$$z(t) \triangleq \int_0^t -\gamma e^{-\gamma \sigma(t-\tau)} \mathcal{B}^\mathrm{T}(\mathfrak{G}) x(\tau) \mathrm{d}\tau, \tag{9}$$

and use $\mathcal{L}(\mathfrak{G}) = \mathcal{B}(\mathfrak{G}) \mathcal{B}(\mathfrak{G})^\mathrm{T}$ to write (5) and (6) as

$$\dot{x}(t) = a_0 x(t) - \alpha \mathcal{F}(\mathfrak{G}) x(t) + \mathcal{B}(\mathfrak{G}) z(t) - \alpha K_1(t) x(t)$$
$$\qquad + \alpha K_2(t) c(t), \tag{10}$$
$$\dot{z}(t) = -\gamma \mathcal{B}^\mathrm{T}(\mathfrak{G}) x(t) - \gamma \sigma z(t), \quad z(0) = \mathbf{0}_N, \tag{11}$$

where $\mathcal{F}(\mathfrak{G}) \triangleq \mathcal{L}(\mathfrak{G}) + \beta \in \mathbb{R}_+^{N\times N}$ based on Lemma 1. It now follows from (8) that

$$\dot{\delta}(t) = a_0 \delta(t) - \alpha \mathcal{F}(\mathfrak{G}) \delta(t) - \alpha K_1(t) \delta(t) + \mathcal{B}(\mathfrak{G}) z(t)$$
$$\qquad + (a_0 \mathrm{I}_N - \alpha \beta) \mathbf{1}_N \epsilon(t) + \alpha L_c(t) K_2(t) c(t) - \mathbf{1}_N \dot{\epsilon}(t),$$
$$\qquad \delta(0) = \delta_0, \tag{12}$$

where $L_c(t) \triangleq \frac{K_1(t) \mathbf{1}_N \mathbf{1}_N^\mathrm{T}}{\mathbf{1}_N^\mathrm{T} K_2(t) \mathbf{1}_N} - \mathrm{I}_N$. We now define the error signal

$$e(t) \triangleq z(t) - \alpha K_c(t) c(t), \tag{13}$$

where $K_c \triangleq \mathcal{B}(\mathfrak{G})^\mathrm{T} \mathcal{L}^\dagger(\mathfrak{G}) L_c(t) K_2(t)$ with the dynamics

$$\dot{e}(t) = -\gamma \mathcal{B}^\mathrm{T}(\mathfrak{G}) \delta(t) - \gamma \sigma e(t) - \alpha \gamma \sigma K_c(t) c(t)$$
$$\qquad -\alpha \dot{K}_c c(t) - \alpha K_c \dot{c}(t), \quad e(0) = e_0. \tag{14}$$

Finally, the closed-loop error dynamics can be expressed as

$$\dot{\delta}(t) = -\mathcal{H}(\mathfrak{G}) \delta(t) - \alpha K_1(t) \delta(t) + \mathcal{B}(\mathfrak{G}) e(t) + s_1(t),$$
$$\qquad \delta(0) = \delta_0, \tag{15}$$
$$\dot{e}(t) = -\gamma \mathcal{B}^\mathrm{T}(\mathfrak{G}) \delta(t) - \gamma \sigma e(t) + s_2(t), \quad e(0) = e_0, \tag{16}$$

where $\mathcal{H}(\mathfrak{G}) \triangleq \alpha \mathcal{F}(\mathfrak{G}) - a_0 \mathrm{I}_N \in \mathbb{R}_+^{N\times N}$ and

$$s_1(t) \triangleq (a_0 \mathrm{I}_N - \alpha \beta) \mathbf{1}_N \epsilon(t) - \mathbf{1}_N \dot{\epsilon}(t), \tag{17}$$
$$s_2(t) \triangleq -\alpha \gamma \sigma K_c(t) c(t) - \alpha \dot{K}_c c(t) - \alpha K_c \dot{c}(t). \tag{18}$$

**Theorem 1** ([5], Theorem 1). *The closed-loop error dynamics (15) and (16) are bounded under the distributed control algorithm (2) and (3), and the ultimate upper bound on the error signal $\delta(t)$ satisfies*

$$\begin{aligned}
\|\delta(t)\|^2 &\leq \frac{\overline{\epsilon}\|\beta\|(\alpha^2\|\beta\| + 2N\alpha\overline{\epsilon}_d) + N^2 \overline{\epsilon}_d^2 + |a_0|N\overline{\epsilon}}{\alpha^2 \lambda_{\min}^2(\mathcal{F}(\mathfrak{G}))} \\
&\quad + \frac{\alpha^2}{\gamma^3 \sigma^2}(\gamma \sigma \overline{p}_1 + \overline{p}_2)^2,
\end{aligned} \tag{19}$$

*where $\|\mathcal{B}^\mathrm{T}(\mathfrak{G})\mathcal{L}^\dagger(\mathfrak{G})K_c(t)c(t)\| \leq \overline{p}_1$ and $\|\mathcal{B}^\mathrm{T}(\mathfrak{G})\mathcal{L}^\dagger(\mathfrak{G}) \cdot \dot{K}_c(t)c(t) + \mathcal{B}^\mathrm{T}(\mathfrak{G})\mathcal{L}^\dagger(\mathfrak{G})K_c(t)\dot{c}(t)\| \leq \overline{p}_2$.*

Theorem 1 shows that the distributed control algorithm by (2) and (3) drives the state of the agents to a neighborhood of the average of the applied inputs. Moreover, by judiciously selection of the design parameters $\alpha, \gamma, \sigma$ and $\beta_i$, this neighborhood can be user-adjustable and get arbitrarily small. Yet, the control algorithm in (2) and (3) requires agent $i$ to have access to its own *exact* states (i.e., $x_i(t)$) as well as the *exact* states of the neighboring agents (i.e., $x_j(t), i \sim j$). From a practical standpoint, this requirement can be limiting in the presence of sensor and information exchange uncertainties, which can be due to the external disturbances, malicious sensor attacks, sensor failure, sensor bias, or detrimental environmental conditions. Consequently, these uncertainties can significantly deteriorate the closed-loop system performance in reaching to the average of the applied inputs in active-passive multiagent system. Next section presents an adaptive control strategy for suppressing the adverse effects of these uncertainties in active-passive multiagent systems.

## IV. Adaptive Active-Passive Control Architecture

In this section, we propose an adaptive control architecture for active-passive multiagent systems subject to information exchange uncertainties. Specifically, consider that each agent has only access to a corrupted measurement of its own state and states of the neighboring agents. Mathematically speaking, the uncertain state measurement of agent $i$ satisfies

$$\tilde{x}_i(t) = x_i(t) + \Delta_i(t), \tag{20}$$

where $\Delta_i(t) \in \mathbb{R}$ is the time-varying uncertainty of agent $i$. For the well-posedness of the considered problem, we now make the following assumption.

**Assumption 2.** *The time-varying uncertainty $\Delta_i(t) \in \mathbb{R}$, $i = 1,\ldots,N$ and its time-derivative are bounded; that is,*

$$|\Delta_i(t)| \leq \overline{\Delta}_i, \quad |\dot{\Delta}_i(t)| \leq \overline{\Delta}_{id}, \tag{21}$$

*where $\overline{\Delta}_i, \overline{\Delta}_{id} \in \mathbb{R}_+$.*

Now, consider the distributed adaptive control architecture given by

$$u_i(t) = -k_0 \tilde{x}_i(t) - \alpha \Big[\sum_{i \sim j}\big(\tilde{x}_i(t) - \tilde{x}_j(t)\big) + \beta_i \tilde{x}_i(t)\Big]$$
$$\qquad + p_i(t) - \alpha \sum_{i \sim h} k_{ih}(t)\big(\tilde{x}_i(t) - c_h(t)\big) + v_i(t), \tag{22}$$

$$\dot{p}_i(t) = -\gamma\Big[\sum_{i\sim j}\big(\tilde{x}_i(t)-\tilde{x}_j(t)\big)+\sigma p_i(t)\Big]+w_i(t),$$
$$p_i(0)=0. \quad (23)$$

In (22) and (23), $v_i(t), w_i(t) \in \mathbb{R}$ are adaptive corrective signals for agent $i$ satisfying

$$v_i(t) \triangleq k_0\hat{\Delta}_i(t)+\alpha\Big[\sum_{i\sim j}\big(\hat{\Delta}_i(t)-\hat{\Delta}_j(t)\big)+\beta_i\hat{\Delta}_i(t)\Big]$$
$$-\alpha\sum_{i\sim h}k_{ih}(t)\hat{\Delta}_i(t), \quad (24)$$
$$w_i(t) \triangleq \gamma\Big[\sum_{i\sim j}\big(\hat{\Delta}_i(t)-\hat{\Delta}_j(t)\big)\Big], \quad (25)$$

where $\hat{\Delta}_i(t)\in\mathbb{R}$ is the estimation of the uncertainty $\Delta_i(t)$ with the update law

$$\dot{\hat{\Delta}}_i(t) = \Gamma\text{Proj}\big(\hat{\Delta}_i(t), -a(\tilde{x}_i(t)-\hat{x}_i(t)-\hat{\Delta}_i(t))\big),$$
$$\hat{\Delta}_i(0)=\hat{\Delta}_{i0}. \quad (26)$$

Here, $\Gamma \in \mathbb{R}_+$ is the adaptation rate and $\hat{\Delta}_{\max} \in \mathbb{R}_+$ is the projection operator bound. In (26), $\hat{x}_i(t)\in \mathbb{R}$ is the estimation of the state of agent $i$ with the update law

$$\dot{\hat{x}}_i(t) = a_0\hat{x}_i(t)-\alpha\Big[\sum_{i\sim j}\big(\hat{x}_i(t)-\hat{x}_j(t)\big)+\beta_i\hat{x}_i(t)\Big]$$
$$+\hat{p}_i(t)-\alpha\sum_{i\sim h}k_{ih}(t)\big(\hat{x}_i(t)-c_h(t)\big)$$
$$+(\Gamma a+\mu)(\tilde{x}_i(t)-\hat{x}_i(t)-\hat{\Delta}_i(t)), \quad \hat{x}_i(0)=\hat{x}_{i0}, \quad (27)$$

where $\hat{p}_i(t) \in \mathbb{R}$ is the estimation of the true integral action of agent $i$ (i.e., the integral action signal when the states are available) having the update law

$$\dot{\hat{p}}_i(t) = -\gamma\Big[\sum_{i\sim j}\big(\hat{x}_i(t)-\hat{x}_j(t)\big)+\sigma\hat{p}_i(t)\Big]$$
$$+\gamma\Big[\sum_{i\sim j}\big(\hat{\Delta}_i(t)-\hat{\Delta}_j(t)\big)\Big], \quad \hat{p}_i(0)=0. \quad (28)$$

By introducing (22) in (1), one can write the system dynamics in the compact form

$$\dot{x}(t) = a_0 x(t)-\alpha\mathcal{F}(\mathfrak{G})x(t)-\alpha K_1(t)x(t)$$
$$+\alpha K_2(t)c(t)+p(t)-k_0\Delta(t)-\alpha\mathcal{F}(\mathfrak{G})\Delta(t)$$
$$-\alpha K_1(t)\Delta(t)+v(t), \quad x(0)=x_0, \quad (29)$$
$$\dot{p}(t) = -\gamma\mathcal{L}(\mathfrak{G})x(t)-\gamma\sigma p(t)-\gamma\mathcal{L}(\mathfrak{G})\Delta(t)+w(t),$$
$$p(0)=\mathbf{0}_N, \quad (30)$$

where $\Delta(t) \triangleq [\Delta_1(t), \Delta_2(t), \ldots, \Delta_N(t)]^\mathrm{T} \in \mathbb{R}^N$, $v(t) \triangleq [v_1(t), v_2(t), \ldots, v_N(t)]^\mathrm{T} \in \mathbb{R}^N$ and $w(t) \triangleq [w_1(t), w_2(t), \ldots, w_N(t)]^\mathrm{T} \in \mathbb{R}^N$. As a direct consequence of the Assumption 2, note that $|\Delta(t)|\leq\overline{\Delta}$, $|\dot{\Delta}(t)|\leq\overline{\Delta}_d$ hold for some $\overline{\Delta}, \overline{\Delta}_d \in \mathbb{R}_+$. Now let

$$z(t) \triangleq \int_0^t -\gamma e^{-\gamma\sigma(t-\tau)}\mathcal{B}^\mathrm{T}(\mathfrak{G})\big(\tilde{x}(\tau)-\gamma^{-1}\mathcal{L}^\dagger(\mathfrak{G})w(\tau)\big)\mathrm{d}\tau,$$
$$(31)$$

where $\tilde{x}(t) \triangleq [\tilde{x}_1(t), \tilde{x}_2(t), \ldots, \tilde{x}_N(t)]^\mathrm{T} \in \mathbb{R}^N$. Using (29) and (31), the uncertain state dynamics can be written as

$$\dot{\tilde{x}}(t) = -\mathcal{H}(\mathfrak{G})\tilde{x}(t)-\alpha K_1(t)\tilde{x}(t)+\mathcal{B}(\mathfrak{G})z(t)$$
$$+\alpha K_2(t)c(t)-a\Delta(t)+\dot{\Delta}(t)+v(t), \quad \tilde{x}(0)=\tilde{x}_0,$$
$$(32)$$
$$\dot{z}(t) = -\gamma\mathcal{B}^\mathrm{T}(\mathfrak{G})\tilde{x}(t)-\gamma\sigma z(t)+\mathcal{B}^\mathrm{T}(\mathfrak{G})\mathcal{L}^\dagger(\mathfrak{G})w(t),$$
$$z(0)=\mathbf{0}_N. \quad (33)$$

It follows from (8) that

$$\dot{\delta}(t) = \dot{\tilde{x}}(t)-\mathbf{1}_N\dot{\epsilon}(t)-\dot{\Delta}(t)$$
$$= -\mathcal{H}(\mathfrak{G})\delta(t)-\alpha K_1(t)\delta(t)+\mathcal{B}(\mathfrak{G})z(t)$$
$$+(a_0-\alpha\beta)\mathbf{1}_N\epsilon(t)-\mathbf{1}_N\dot{\epsilon}(t)-\alpha L_c K_2(t)c(t)$$
$$-(a\mathrm{I}_N+H+\alpha K_1(t))\Delta(t)+v(t),$$
$$= -\mathcal{H}(\mathfrak{G})\delta(t)-\alpha K_1(t)\delta(t)+\mathcal{B}(\mathfrak{G})e(t)$$
$$+(a_0-\alpha\beta)\mathbf{1}_N\epsilon(t)-\mathbf{1}_N\dot{\epsilon}(t)$$
$$-(a\mathrm{I}_N+H+\alpha K_1(t))\Delta(t)+v(t), \quad \delta(0)=\delta_0. \quad (34)$$

From (13), the error dynamics can also be written as

$$\dot{e}(t) = -\gamma\mathcal{B}^\mathrm{T}(\mathfrak{G})\delta(t)-\gamma\sigma e(t)-\alpha\gamma\sigma K_c(t)c(t)$$
$$-\mathcal{B}^\mathrm{T}(\mathfrak{G})\big(\gamma\Delta(t)-\mathcal{L}^\dagger(\mathfrak{G})w(t)\big)-\alpha\dot{K}_c c(t)$$
$$-\alpha K_c \dot{c}(t), \quad e(0)=e_0. \quad (35)$$

We now substitute the corrective signals from (24) and (25) into (34) and (35), which gives the closed-loop error dynamics

$$\dot{\delta}(t) = -\mathcal{H}(\mathfrak{G})\delta(t)-\alpha K_1(t)\delta(t)+\mathcal{B}(\mathfrak{G})e(t)-(\alpha\mathcal{F}(\mathfrak{G})$$
$$+\alpha K_1(t)+k_0\mathrm{I}_N)\tilde{\Delta}(t)+s_1(t), \quad \delta(0)=\delta_0, \quad (36)$$
$$\dot{e}(t) = -\gamma\mathcal{B}^\mathrm{T}(\mathfrak{G})\delta(t)-\gamma\sigma e(t)-\gamma\mathcal{B}^\mathrm{T}(\mathfrak{G})\tilde{\Delta}(t)+s_2(t),$$
$$e(0)=e_0, \quad (37)$$

where $s_1(t)$ and $s_2(t)$ are given in (17) and (18), and $\tilde{\Delta}(t) \triangleq \Delta(t)-\hat{\Delta}(t) \in \mathbb{R}^N$ is the uncertainty estimation error.

Next, define $e_{x_i}(t) \triangleq \tilde{x}_i(t)-\hat{x}_i(t)-\hat{\Delta}_i(t) \in \mathbb{R}$. Note that the compact form of (27) can be written as

$$\dot{\hat{x}}(t) = -\mathcal{H}(\mathfrak{G})\hat{x}(t)-\alpha K_1(t)\hat{x}(t)+\alpha K_2(t)c(t)$$
$$+(-\alpha\mathcal{F}(\mathfrak{G})-\alpha K_1(+\mathcal{B}(\mathfrak{G})\hat{z}(t)t)-k_0\mathrm{I}_N)\hat{\Delta}(t)$$
$$+v(t)-\dot{\hat{\Delta}}(t)+\mu e_x(t), \quad \hat{x}(0)=\hat{x}_0, \quad (38)$$

where $e_x(t) \triangleq [e_{x_1}(t), e_{x_2}(t), \ldots, e_{x_N}(t)]^\mathrm{T} \in \mathbb{R}^N$ and $\hat{z}(t)\in\mathbb{R}^N$ is the estimation of $z(t)$ in (31) given by

$$\hat{z}(t) \triangleq \int_0^t -\gamma e^{-\gamma\sigma(t-\tau)}\mathcal{B}^\mathrm{T}(\mathfrak{G})\hat{x}(\tau)\mathrm{d}\tau. \quad (39)$$

Defining $e_z(t) \triangleq z(t)-\hat{z}(t) \in \mathbb{R}^N$, one can write the estimation error dynamics for $e_x(t), e_z(t)$ and $\tilde{\Delta}(t)$ respectively as

$$\dot{e}_x(t) = -(\mathcal{H}(\mathfrak{G})+\mu\mathrm{I}_N)e_x(t)-\alpha K_1(t)e_x(t)-a\tilde{\Delta}(t)$$
$$+\mathcal{B}(\mathfrak{G})e_z(t)+\dot{\Delta}(t), \quad e_x(0)=e_{x0}, \quad (40)$$
$$\dot{e}_z(t) = -\gamma\sigma e_z(t)-\gamma\mathcal{B}^\mathrm{T}(\mathfrak{G})e_x(t), \quad e_z(0)=e_{z0}, \quad (41)$$
$$\dot{\tilde{\Delta}}(t) = \dot{\Delta}(t)-\Gamma\text{Proj}\big(\hat{\Delta}(t), -ae_x(t)\big), \quad \tilde{\Delta}(0)=\tilde{\Delta}_0. \quad (42)$$

**Theorem 2.** *Consider an active-passive multiagent system consisting of $N$ agents over a connected, undirected graph $\mathfrak{G}$. In addition, consider the dynamics of each agent given by (1), where the state measurements are corrupted by the unknown time-varying uncertainty $\Delta_i(t)$ predicated on (20). Let the adaptive distributed control signal be given by (22) and (23) along with the corrective signals (24) and (25) and the update law (26). The closed-loop error dynamics (40), (41) and (42) are then bounded subject to the ultimate bounds given by*

$$\|e_x(t)\| \leq \sqrt{\frac{\gamma_1}{\gamma_0}\eta_1^2 + \frac{\Gamma^{-1}}{\gamma_0}\eta_2^2}, \tag{43}$$

$$\|e_z(t)\| \leq \sqrt{\frac{\gamma_1}{\gamma_0}\eta_1^2 + \frac{\Gamma^{-1}}{\gamma_0}\eta_2^2}, \tag{44}$$

$$\|\tilde{\Delta}(t)\| \leq \sqrt{\gamma_1 \Gamma \eta_1^2 + \eta_2^2}, \tag{45}$$

*where $\gamma_0 \triangleq 0.5\min(1,\gamma^{-1})$, $\gamma_1 \triangleq 0.5\max(1,\gamma^{-1})$, $\eta_1 \triangleq \frac{\alpha_2}{2\alpha_0} + (\frac{\alpha_2^2}{4\alpha_0^2} + \frac{2\Gamma^{-1}\alpha_2\eta_2}{\alpha_0})^{1/2}$, $\eta_2 \triangleq \overline{\Delta} + \hat{\Delta}_{\max}$, $\alpha_1 \triangleq \lambda_{\min}(\mathcal{H}(\mathfrak{G}) + \mu I_N) \in \mathbb{R}_+$, $\alpha_0 \triangleq \min(\alpha_1,\sigma) \in \mathbb{R}_+$, $\alpha_2 \triangleq \overline{\Delta}_d \in \overline{\mathbb{R}}_+$.*

*Proof.* Consider the Lyapunov candidate function

$$V(e_x, e_z, \tilde{\Delta}) = \sum_{i=1}^{N} \frac{1}{2}\|e_{xi}\|^2 + \frac{\gamma^{-1}}{2}\|e_{zi}\|^2 + \Gamma^{-1}\|\tilde{\Delta}_i\|^2,$$

$$= \frac{1}{2}e_x^T e_x + \frac{\gamma^{-1}}{2}e_z^T e_z + \Gamma^{-1}\tilde{\Delta}^T\tilde{\Delta}. \tag{46}$$

The time derivative of (46) along the closed-loop system trajectories (40), (41) and (42) is given by

$$\begin{aligned}
\dot{V}(\cdot) &= -e_x^T(t)(\mathcal{H}(\mathfrak{G})+\mu I_N)e_x(t) - \alpha e_x^T(t)K_1(t)e_x(t) \\
&\quad -ae_x^T(t)\tilde{\Delta}(t) + e_x^T(t)\mathcal{B}(\mathfrak{G})e_z(t) + e_x^T(t)\dot{\Delta}(t) \\
&\quad -\sigma e_z^T(t)e_z(t) - e_z^T(t)\mathcal{B}^T(\mathfrak{G})e_x(t) \\
&\quad +2\Gamma^{-1}\tilde{\Delta}^T(t)\dot{\tilde{\Delta}}(t), \\
&\leq -e_x^T(t)(\mathcal{H}(\mathfrak{G})+\mu I_N)e_x(t) - ae_x^T(t)\tilde{\Delta}(t) \\
&\quad +e_x^T(t)\dot{\Delta}(t) - \sigma e_z^T(t)e_z(t) + 2\Gamma^{-1}\tilde{\Delta}^T(t)\dot{\tilde{\Delta}}(t), \\
&\leq -e_x^T(t)(\mathcal{H}(\mathfrak{G})+\mu I_N)e_x(t) - \sigma e_z^T(t)e_z(t) \\
&\quad +2(\hat{\Delta}(t)-\Delta(t))^T\big(\text{Proj}(\hat{\Delta}(t),-ae_x(t)) + ae_x(t)\big) \\
&\quad +2\Gamma^{-1}\tilde{\Delta}^T(t)\dot{\Delta}(t) + e_x^T(t)\dot{\Delta}(t). \tag{47}
\end{aligned}$$

Applying the projection operator property as stated after Definition 1 yields

$$\begin{aligned}
\dot{V}(\cdot) &\leq -e_x^T(t)(\mathcal{H}(\mathfrak{G})+\mu I_N)e_x(t) - \sigma e_z^T(t)e_z(t) \\
&\quad +2\Gamma^{-1}\tilde{\Delta}^T(t)\dot{\Delta}(t) + e_x^T(t)\dot{\Delta}(t). \tag{48}
\end{aligned}$$

The upper bound of (48) can now be written as

$$\begin{aligned}
\dot{V}(\cdot) &\leq -\alpha_1\|e_x(t)\|^2 + \alpha_2\|e_x(t)\| - \sigma\|e_z(t)\| \\
&\quad +2\Gamma^{-1}\alpha_2\eta_2. \tag{49}
\end{aligned}$$

Next, define $e_s(t) \triangleq [e_x(t)^T e_z(t)^T]^T \in \mathbb{R}^{2N}$. One can then rewrite (49) as

$$\dot{V}(\cdot) \leq -\alpha_0\|e_s(t)\|^2 + \alpha_2\|e_s(t)\| + 2\Gamma^{-1}\alpha_2\eta_2,$$

$$\leq -\big[\sqrt{\alpha_0}\|e_s(t)\| - \frac{\alpha_2}{2\sqrt{\alpha_0}}\big]^2 + \frac{\alpha_2^2}{4\alpha_0} + 2\Gamma^{-1}\alpha_2\eta_2. \tag{50}$$

It follows from (50) that $\dot{V}(\cdot) < 0$ outside the compact set $\Omega \triangleq \{(e_s(t),\tilde{\Delta}(t)) : \|e_s(t)\| \leq \eta_1, \|\tilde{\Delta}(t)\| \leq \eta_2\}$, where this guarantees the uniform boundedness of the solution $(e_x(t), e_z(t), \tilde{\Delta}(t))$ of the closed-loop system (40), (41) and (42). Finally, to calculate the corresponding ultimate bounds, one can write (46) as

$$V(\cdot) = e_s^T(t)\begin{bmatrix} \frac{1}{2} & 0 \\ 0 & \frac{\gamma^{-1}}{2} \end{bmatrix}e_s(t) + \Gamma^{-1}\tilde{\Delta}^T(t)\tilde{\Delta}(t). \tag{51}$$

Hence $\gamma_0\|e_s(t)\|^2 + \Gamma^{-1}\|\tilde{\Delta}(t)\| \leq \gamma_1\eta_1^2 + \gamma_2\eta_2^2$ holds after the transient time, that is equivalent to $\gamma_0\|e_x(t)\|^2 \leq \gamma_1\eta_1^2 + \gamma_2\eta_2^2$, $\gamma_0\|e_z(t)\|^2 \leq \gamma_1\eta_1^2 + \gamma_2\eta_2^2$, and $\Gamma^{-1}\|\tilde{\Delta}(t)\| \leq \gamma_1\eta_1^2 + \gamma_2\eta_2^2$, which gives the ultimate bounds in (43)-(45). ∎

In the next corollary, we present a special case when the communication uncertainties are constant.

**Corollary 1.** *Consider the active-passive multiagent system as described in Theorem 2. In addition, consider that the communication uncertainties are time-invariant; that is, $\Delta_i(t) \equiv \Delta_i$ in (20). Let the adaptive distributed control signal be given by (22) and (23) along with the corrective signals (24) and (25) and the update law*

$$\dot{\hat{\Delta}}_i(t) = -\Gamma a(\tilde{x}_i(t) - \hat{x}_i(t) - \hat{\Delta}_i(t)). \quad \hat{\Delta}_i(0) = \hat{\Delta}_{i0}, \tag{52}$$

*The closed-loop error dynamics (40), (41) and (42) are then bounded and $\lim_{t\to\infty} e_x(t) = \lim_{t\to\infty} e_z(t) = \lim_{t\to\infty} \tilde{\Delta}(t) = 0$.*

*Proof.* The result directly follows from the proof of Theorem 2; hence, the proof is omitted. ∎

**Remark 2.** *Since the estimation error for the uncertainties $\tilde{\delta}(t)$ is shown to be bounded in Theorem 2, one can write (36) and (37) in the form of (15) and (16). Consequently, similar ultimate upper bound for the error signal $\delta(t)$ can be calculated based on Theorem 1 for the case of uncertain information exchange. Moreover, for the case of time-invariant uncertainties, the estimation of the uncertainties $\hat{\Delta}(t)$ converge to $\Delta$ based on Corollary 1, and the estimation of the system state $\hat{x}(t)$ converges to $x(t)$. This recovers the original performance of the active-passive multiagent system as stated in Theorem 1.*

## V. ILLUSTRATIVE NUMERICAL EXAMPLE

In this section, we consider a coverage control problem predicated on the proposed adaptive active-passive multiagent system framework to illustrate the efficacy of the proposed approach. In particular, consider an orchard as shown in Figure 2, where a farmer needs to visit several task locations. We assume that a group of mobile sensors are spread throughout the field and are required to assist the human at the task location that are not known for the agents a-priori. Consider that the agents can locally sense the position and the velocity of the human when the human is within their sensing radius $R_s$. The mobile sensors then infer

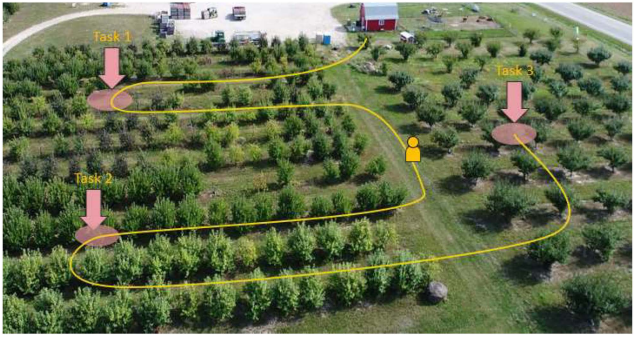

Fig. 2. A human visiting three task locations in an orchard (Image taken from `https://www.tripadvisor.com/`, Overhiser Orchards, South Haven, MI).

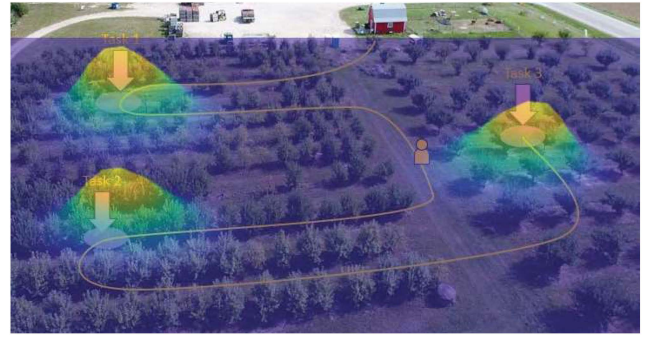

Fig. 3. An illustration of the density map for the task locations.

the task locations based on the estimated locations where the human slows down or stops.

For this purpose, the proposed adaptive active-passive control architecture is used to estimate the position and the velocity of the human in a distributed manner. Each agent then constructs a density map based on the estimated human position and velocity such that a lower human speed corresponds to a higher density function value. As a result, the obtained time-varying density map represents an *importance* map for the task locations (see Figure 3). We then utilize the coverage control method introduced in [21] to coordinate the agents based on the time-varying density map toward the task locations.

For this simulation study, we consider 25 agents with the sensing dynamics given in (1) with $a = 1$. The agents are considered to be initialized at equal distance from each other; that is, they cover a 20 by 20 meters field $D = \{q \mid q \in [0, 20] \times [0, 20]\}$, and they are connected based on an connected, undirected graph similar to the graph shown in Figure 1. In addition, the position of the mobile sensors satisfy the dynamics given by

$$\dot{O}_i(t) = U_i(t), \quad i = 1, \ldots, N, \quad (53)$$

where $O_i(t) = [O_{xi}(t), O_{yi}(t)]^\mathrm{T} \in \mathbb{R}^2$ denotes the position of agent $i$ in the two-dimensional space and $U_i(t) = [U_{xi}(t), U_{yi}(t)]^\mathrm{T} \in \mathcal{U} \subset \mathbb{R}^2$ denotes the corresponding control signal. The human position and velocity data are simulated in MATLAB by moving the mouse pointer within a graph representing the orchard by the user. The proposed adaptive active-passive architecture in Section IV is used for estimating the $x$ and $y$ components of the position of the human (i.e., $x_x(t)$ and $x_y(t)$) as well as the magnitude of its velocity $v$ (i.e., $x_v(t)$). In what follows, the subscripts $x, y$ and $v$ denotes each design parameter associated with estimation of the $x$ and $y$ position and the velocity of the human, respectively. We set $k_0 = 1, \gamma_x = \gamma_y = 22, \gamma_v = 30$, $\sigma_x = \sigma_y = 0.0045, \sigma_v = 0.0033, \alpha_x = \alpha_y = 20, \alpha_v = 30$, $\beta_i = 0.001, i = 1, \ldots, 25$ $\mu = 1.5, \Gamma_x = \Gamma_y = 5, \Gamma_v = 8$ and $R_s = 3.5 \ m$. The information exchange uncertainties for position, $\Delta_{xi}$ and $\Delta_{yi}$, and velocity $\Delta_{vi}$, are selected randomly within the intervals $[0, 5] \ m$ and $[0, 1] \ m/s$, respectively. Figures 4 and 5 show the estimation of the position and velocity of the human by the active-passive multiagent system. The initial deviation in position estimation during the transient period of the parameter adjustment mechanism can be seen around $(0, 0)$ position in Figure 4.

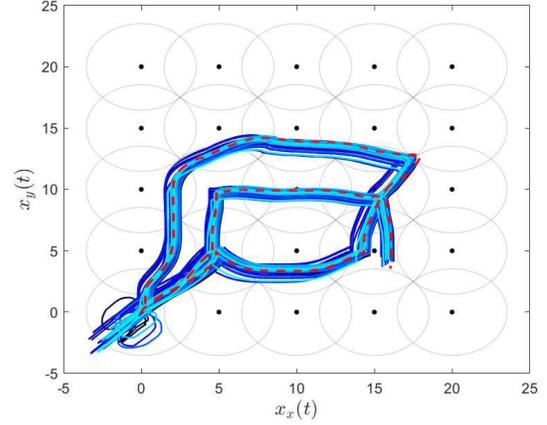

Fig. 4. Estimation of the position of the human by the active-passive network. The red dashed line denotes the trajectory of the human, the solid blue lines are the estimated trajectories, and the black dots denote the initial position of the agents.

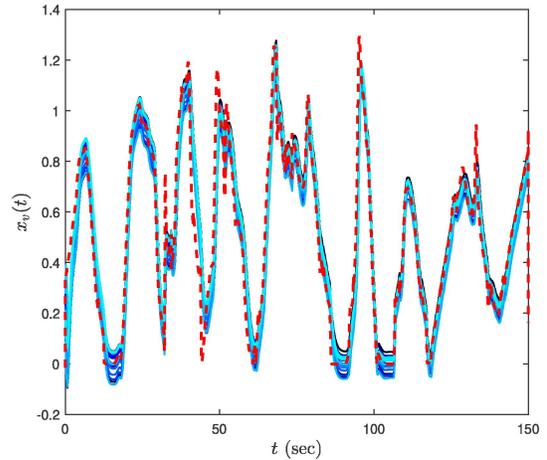

Fig. 5. Estimation of the velocity of the human by the active-passive network. The red dashed line denotes the velocity of the human and the solid blue lines are the estimated velocities.

For the coordination of the mobile sensors, we make use of the coverage control algorithm in [21]. Specifically, a Voronoi tessellation is considered for each mobile sensor $V_i(O) = \{p \in D \mid \|p - O_i\| \le \|p - O_j\|, \forall (i, j) \in \mathcal{E}_\mathfrak{G}\}$. Each agent uses the position and velocity estimates of the human

to construct a time-varying density map $\phi(q,t): D \to [0,\infty)$ that encodes the relative importance of the point in the two-dimensional space $D$ such that

$$\dot{\phi}(q,t) = \frac{1}{|x_v(t)| + 0.1}. \tag{54}$$

This design for the density function increases the relative importance of the points in $D$ when the estimated human velocity is low, that results in indication of the task locations. We now define the cost function

$$\mathcal{H}(O,t) = \sum_{i=1}^{N} \int_{V_i(O)} \|O_i - q\|^2 \phi(q,t) \mathrm{d}q, \tag{55}$$

that describes the quality of the coverage by the mobile sensors. A necessary condition for minimizing (55) is that the mobile sensors locate at the centroid of their corresponding Voronoi tessellations. To this end, we also define the cost function $J(O,t)$ that has stationary points at the centroid of the Voronoi tessellations given by

$$J(O) = \sum_{i=1}^{N} \frac{1}{2} \|O_i - G_i(O,t)\|^2, \tag{56}$$

where $G_i(O,t)$ denotes the center of mass of the Voronoi cell of agent $i$ given by

$$G_i(O,t) = \frac{\int_{V_i(O)} q\phi(q,t)\mathrm{d}q}{\int_{V_i(O)} \phi(q,t)\mathrm{d}q}. \tag{57}$$

The control signal $U_i$ for coordination of sensor $i$ is then calculated by solving the optimization problem [21]

$$\min_{U_i, r_i} \|U_i\|^2 + |r_i|^2$$
$$\text{s.t.} \quad -(O_i - G_i(O,t))^\mathrm{T} \big(I - \frac{\partial G_i(O,t)}{\partial O_i}\big) U_i$$
$$\geq -\alpha(-J_i(O,t)) - (O_i - G_i(O,t))^\mathrm{T} \frac{\partial G_i(O,t)}{\partial t} - r_i, \tag{58}$$

where $\alpha$ is an extended class $\mathcal{K}$ function and $r_i$ is a slack variable. In this simulation study, we assume that the connectivity of the network is preserved. However, one may resort to the connectivity maintenance algorithms, such as [22] to include additional constraints for connectivity to the optimization problem.

Figure 6 shows the coordination of the mobile sensors based on the constructed density map. Note that, the density map shown in this figure is for agent 1 (initially located as $(0,0)$). However, since the estimated human position and velocity for all agents are close to each other based on Figures 4 and 5, the density maps of the other mobile sensors are similar.

## VI. Conclusion

In this paper, we proposed an adaptive control architecture for active-passive multiagent systems that are subject to information exchange uncertainties. A distributed algorithm was designed to estimation these uncertainties, and to drive the agents to a user-adjustable neighborhood of the average of the applied inputs to the active agents. We illustrated the efficacy of the proposed approach in a coverage control problem to mitigate the adverse effects of information exchange uncertainties within a mobile sensor network.

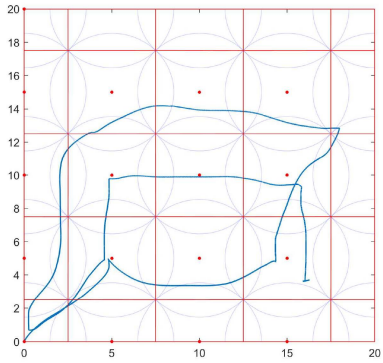 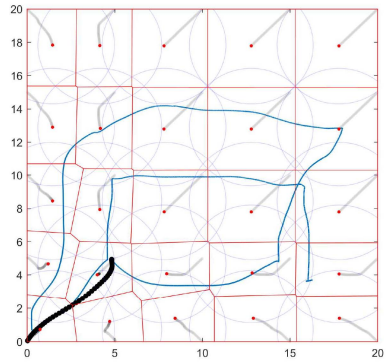 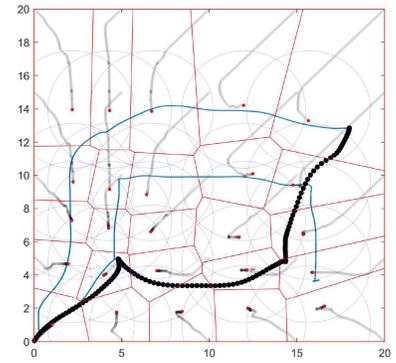
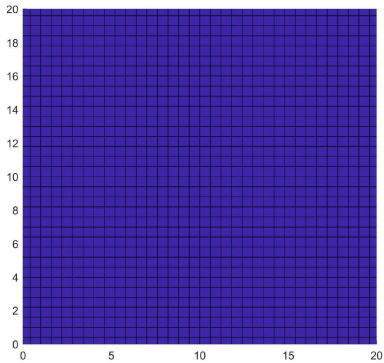 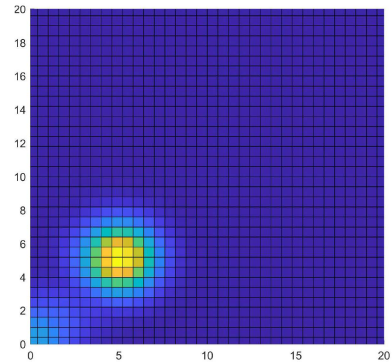 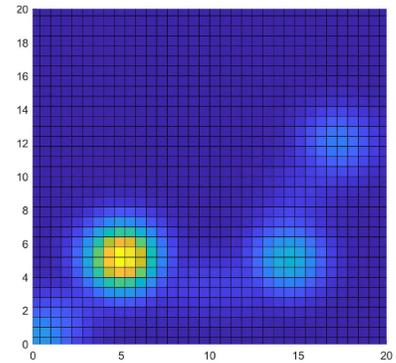

(a) $t = 0$ sec        (b) $t = 15$ sec        (c) $t = 45$ sec

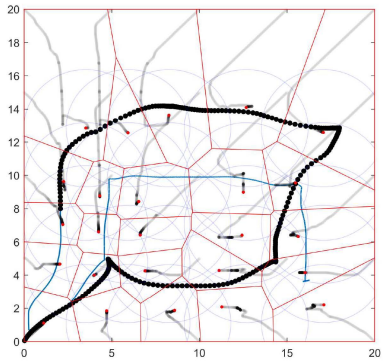 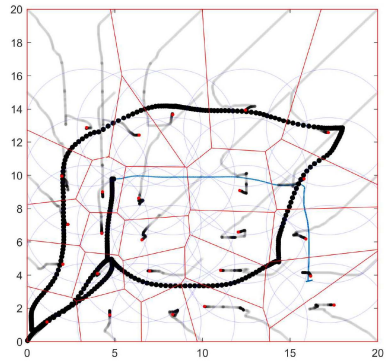 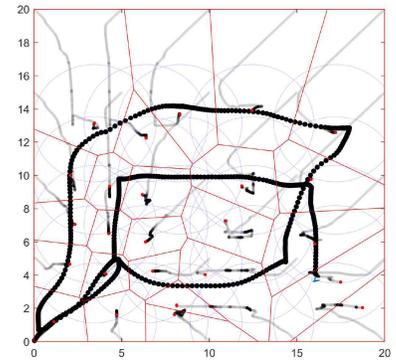
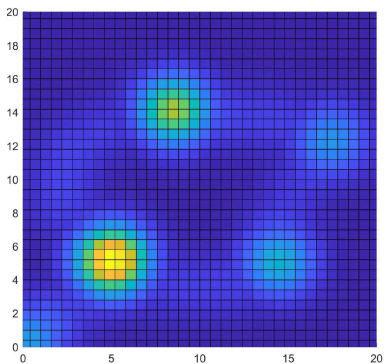 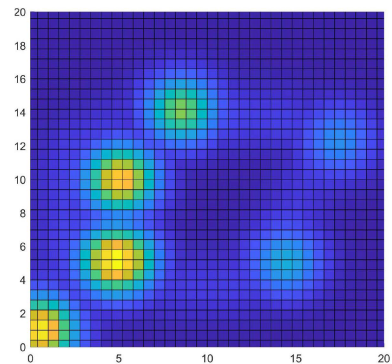 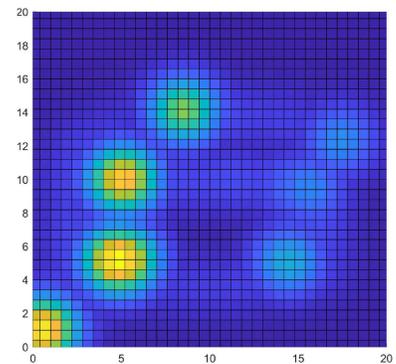

(d) $t = 75$ sec        (e) $t = 120$ sec        (f) $t = 150$ sec

Fig. 6. The evolution of the active-passive multiagent system and the density function over time. The mobile sensors are shown by red dots and the black doted line denotes the actual trajectory of the human.